\documentstyle[epsfig,amssymb,floatflt,float,here]{mn}
\newif\ifAMStwofonts

\newcommand{\sm}{\, {\rm M}_{\odot}}
\newcommand{\sL}{\, {\rm L}_{\odot}}
\newcommand{\kms}{\,{\rm km \, s^{-1}}}

\renewcommand{\d}{{\rm d}}
\newcommand{\e}{{\rm e}}

\ifoldfss
  \ifCUPmtlplainloaded \else
    \NewTextAlphabet{textbfit} {cmbxti10} {}
    \NewTextAlphabet{textbfss} {cmssbx10} {}
    \NewMathAlphabet{mathbfit} {cmbxti10} {} 
    \NewMathAlphabet{mathbfss} {cmssbx10} {} 
  \fi
  \ifAMStwofonts
    \ifCUPmtlplainloaded \else
      \NewSymbolFont{upmath} {eurm10}
      \NewSymbolFont{AMSa} {msam10}
      \NewMathSymbol{\upi}     {0}{upmath}{19}
      \NewMathSymbol{\umu}     {0}{upmath}{16}
      \NewMathSymbol{\upartial}{0}{upmath}{40}
      \NewMathSymbol{\leqslant}{3}{AMSa}{36}
      \NewMathSymbol{\geqslant}{3}{AMSa}{3E}

    \fi
  \fi
\fi 

\ifnfssone
  \newmathalphabet{\mathit}
  \addtoversion{normal}{\mathit}{cmr}{m}{it}
  \addtoversion{bold}{\mathit}{cmr}{bx}{it}
  \newmathalphabet{\mathbfit} 
  \addtoversion{normal}{\mathbfit}{cmr}{bx}{it}
  \addtoversion{bold}{\mathbfit}{cmr}{bx}{it}
  \newmathalphabet{\mathbfss} 
  \addtoversion{normal}{\mathbfss}{cmss}{bx}{n}
  \addtoversion{bold}{\mathbfss}{cmss}{bx}{n}
  \ifAMStwofonts
    \ifCUPmtlplainloaded \else
      \UseAMStwoboldmath
      \makeatletter
      \new@mathgroup\upmath@group
      \define@mathgroup\mv@normal\upmath@group{eur}{m}{n}
      \define@mathgroup\mv@bold\upmath@group{eur}{b}{n}
      \edef\UPM{\hexnumber\upmath@group}
      \new@mathgroup\amsa@group
      \define@mathgroup\mv@normal\amsa@group{msa}{m}{n}
      \define@mathgroup\mv@bold\amsa@group{msa}{m}{n}
      \edef\AMSa{\hexnumber\amsa@group}
      \makeatother
      \mathchardef\upi="0\UPM19
      \mathchardef\umu="0\UPM16
      \mathchardef\upartial="0\UPM40
      \mathchardef\leqslant="3\AMSa36
      \mathchardef\geqslant="3\AMSa3E
    \fi
  \fi
\fi 

\ifnfsstwo
  \DeclareMathAlphabet{\mathbfit}{OT1}{cmr}{bx}{it}
  \SetMathAlphabet\mathbfit{bold}{OT1}{cmr}{bx}{it}
  \DeclareMathAlphabet{\mathbfss}{OT1}{cmss}{bx}{n}
  \SetMathAlphabet\mathbfss{bold}{OT1}{cmss}{bx}{n}
  \ifAMStwofonts
    \ifCUPmtlplainloaded \else
      \DeclareSymbolFont{UPM}{U}{eur}{m}{n}
      \SetSymbolFont{UPM}{bold}{U}{eur}{b}{n}
      \DeclareSymbolFont{AMSa}{U}{msa}{m}{n}
      \DeclareMathSymbol{\upi}{0}{UPM}{"19}
      \DeclareMathSymbol{\umu}{0}{UPM}{"16}
      \DeclareMathSymbol{\upartial}{0}{UPM}{"40}
      \DeclareMathSymbol{\leqslant}{3}{AMSa}{"36}
      \DeclareMathSymbol{\geqslant}{3}{AMSa}{"3E}
    \fi
  \fi
\fi 

\ifCUPmtlplainloaded \else
  \ifAMStwofonts \else 
    \def\upi{\pi}
    \def\umu{\mu}
    \def\upartial{\partial}
  \fi
\fi

\title[Mapping the substructure in the Galactic halo]{Mapping the substructure in the Galactic halo \\ with the next 
generation of astrometric satellites}
\author[A. Helmi and P.T. de Zeeuw]
       {Amina Helmi and P. Tim de Zeeuw\\
        Sterrewacht Leiden, Postbus 9513, 2300 RA Leiden, 
The Netherlands}
\date{Accepted ...
      Received ...;
      in original form ...}

\pagerange{\pageref{firstpage}--\pageref{lastpage}}
\pubyear{}

\begin{document}

\maketitle

\label{firstpage}

\begin{abstract}
 We run numerical simulations of the disruption of satellite galaxies
in a Galactic potential to build up the entire stellar halo, in order to
investigate what the next generation of astrometric satellites will
reveal by observing the halo of the Milky Way.  We generate artificial
DIVA, FAME and GAIA halo catalogues, in which we look for the
signatures left by the accreted satellites. We develop a method based
on the standard Friends-of-Friends algorithm applied to the space of
integrals of motion. We find this simple method can recover about 50\%
of the different accretion events, when the observational
uncertainties expected for GAIA are taken into account, even when the
exact form of the Galactic potential is unknown. The recovery
rate for DIVA and FAME is much smaller, but these missions, like GAIA,
should be able to test the hierarchical formation paradigm on our
Galaxy by measuring the amount of halo substructure in the form of
nearby kinematically cold streams with for example, a two--point correlation
function in velocity space.
\end{abstract}


\section{Introduction}

Hierarchical theories of structure formation in the Universe propose
that galaxies are the result of mergers and accretion of smaller
building blocks (White \& Rees 1978). Detailed studies of the
properties of a galaxy built in this way have shown that such events
leave fossil signatures in the present day components, which for a
galaxy like our own would be clearly detectable with future
astrometric missions (Helmi \& White 1999).  In particular the stellar
halo would be the natural place to look for such substructures, since
a spheroidal component is formed by the trails of stars left by
disrupted satellite galaxies. Moreover, recent observations have shown
that indeed considerable structure is still present in Milky Way's
halo, indicating that accretion events have had some role in its
formation history (e.g.  Ibata, Gilmore \& Irwin 1994; Majewski, Munn
\& Hawley 1996; Helmi et al. 1999; Ivezic et al. 2000).

In the next ten years, several satellite missions will be devoted to
measure with very high accuracy the motions of thousands to many
millions of stars in our Galaxy. NASA's Space Interferometry Mission
(SIM) is a targeted mission which will obtain parallaxes and proper
motions for about 10000 stars. With somewhat different goals, and
more similar to the HIPPARCOS satellite, the Full-sky Astrometric
Mapping Explorer (FAME, Horner et al. 1999) promises to measure
positions and parallaxes for stars brighter than $V \sim 9$ to better
than 50 $\mu$as and proper motions to 50 $\mu$as yr$^{-1}$. At $V\sim
15$ these accuracies will be degraded by an order of magnitude. The
resulting astrometric database will have $4 \times 10^7$ stars, and
may be combined with the radial velocities from the Sloan Digital Sky
Survey or from other ground based catalogues to obtain full
phase-space information. Less ambitious but still an improvement over
HIPPARCOS is the German DIVA mission (R\"oser 1998). If launched it
will observe of the order of $3.5 \times 10^7$ stars, at four times the
precision of HIPPARCOS ($\sigma_{\pi} = 0.25$ mas and $\sigma_{\mu} =
0.4$ mas yr$^{-1}$ at $V = 10$), thereby completing the knowledge of
nearby stars. Like FAME, DIVA will not measure radial velocities.  On
the other hand, the proposed ESA astrometric satellite GAIA (Gilmore
et al. 1998) will provide very precise astrometry ($<$10 $\mu$as in
parallax and $<$10 $\mu$as yr$^{-1}$ in proper motion at $V \sim 15$,
increasing to 0.2 mas yr$^{-1}$ at $V\sim 20$) and multicolour
photometry, for all 1.3 billion objects to $V\sim20$, {\it and} radial
velocities with accuracies of a few $\kms$ for most stars brighter
than $V\sim17$, so that full and homogeneous six-dimensional
phase-space information will be available.  These satellite missions
will thereby provide a very large and statistically reliable sample of
stars, from which the fundamental questions concerning the origin and
evolution of the Galaxy may finally be answered. 

In this paper we shall focus on what GAIA will tell us about the
history and formation of the stellar halo of the Milky Way. We will
also discuss the impact of DIVA and FAME, and leave aside SIM as this
mission will not provide a survey but a hand-picked catalogue of
stars. Even though we focus on the stellar halo, the method that we
shall propose for finding substructures in phase-space may also be
extended to find, for example, disk moving groups (e.g. de Zeeuw et
al. 1999; Chereul, Cr\'ez\'e \& Bienaym\'e 1999).

There are several methods for detecting moving groups.  The Great
Circle Cell Counts method (G3C) proposed by Johnston, Hernquist \& Bolte
(1996) uses the position on the sky, and employs the fact that
satellite galaxies in orbits that probe only the outer (spherical) halo
conserve the orientation of their plane of motion, thereby leaving
their debris along great circles on the sky, if observed from the
Galactic centre.  The methods used in
the Solar neighbourhood for detection of disk moving groups and open
clusters use also proper motions (and sometimes parallax), and assume
that all the stars belonging to the same system have the same velocity
vector (e.g.\ Hoogerwerf \& Aguilar 1998; de Bruijne 1998).
Lynden-Bell \& Lynden-Bell's method (1995) needs the position on the
sky and the radial velocity, and has been used, for example, to link
globular clusters which lie in the same plane to some of the
(disrupted) dwarf companions of our Galaxy (see also Lynden-Bell,
1999).  The applicability of the above mentioned methods is
questionable in the inner parts of the halo. In this regime, the
Galactic potential is significantly flattened so that the debris does
not remain on a fixed plane, the situation where G3C works. As noted
by Helmi \& White (1999) no spatial correlations should be expected
for satellites disrupted several Gyr ago. 
On the other hand, even though the velocity
dispersions in a stellar stream do decrease with time, and therefore,
very strong correlations are to be expected, in
the inner halo strong phase-mixing takes place. For example in the 
Solar neighbourhood several hundred (mainly) cold streams originating
in disrupted satellites may be present, but it may in practice be  
difficult to resolve each one of such moving groups completely.
Clearly, before exploring the full
capabilities of the next generation of astrometric satellite missions,
we first need to identify where the clustering that is characteristic
of a satellite manifests itself in the debris that we observe after
many galactic orbits. As shown in Helmi et al. (1999) a
method based on the lumpiness in integrals of motion space seems to
be a promising tool for unveiling the merger history of our Galaxy.

\section{Building up a stellar halo}

Our main goal is to test whether with the next generation of 
astrometric satellites, we would be able to find the signatures
left by merger events in the Galactic stellar halo. 
We will assume that the whole stellar halo is the result of
the superposition of several disrupted satellite galaxies which fell onto the
Milky Way about 10 Gyr ago. We shall here discuss the initial conditions
and the numerical methods used to generate this version of the stellar halo.

\subsection{Initial conditions for the satellites}
\subsubsection{Orbital properties}

The stellar halo has a density profile  (Kinman, Suntzeff \& Kraft 1994)
\begin{equation}
\label{gaia_eq:rho_r}
\rho_\star(r) = \rho_0 \left(\frac{r}{r_0}\right)^{-3.5}, 
\end{equation}
a total luminosity of about $10^9 \sL$, and a half light radius which
probably lies around 3 kpc from the Galactic centre. For $r = r_0 = 8$ kpc
(the distance to the Galactic Centre from the Sun), $\rho_0$ 
corresponds to the local stellar halo density, for which we take
$\rho_0 = 1.5 \times 10^4 \sm$ kpc$^{-3}$ (Fuchs \& Jahrei\ss \,\, 1998).

The initial orbital conditions of our satellites should be drawn from
the Galactic halo distribution function (DF), which we assume to be a
function of energy $E$ and angular momentum $L$: $f(E,L)$. For
simplicity here we shall assume that the stellar halo is a power-law
tracer population embedded in a singular isothermal sphere,
representing the dark-matter halo of the Milky Way. Following van den
Bosch et al. (1998), we assume that
\[f(E,L) = g(E) h_\alpha(\eta), \qquad \mbox{where} \qquad \eta = L/L_c(E),\] 
and $L_c(E)$ is the angular momentum of a circular orbit with energy $E$: 
$L_c(E) = r_c(E) V_c $, where 
$r_c(E) = {\rm e}^{-1/2} \exp\left[E/V_c^2\right]$. The function
$h_\alpha(\eta)$ is known as the circularity, and determines the degree
of anisotropy of the DF. We choose a simple parametrization of 
$h_\alpha(\eta)$ (Gerhard 1991):  
\begin{equation}
h_\alpha(\eta) =\left\{\begin{array}{ll}
 \tanh\left(\frac{\eta}{\alpha}\right)/\tanh\left(\frac{1}{\alpha}\right) 
& \alpha > 0 \\
 1 & \alpha = 0 \\
 \tanh\left(\frac{1 - \eta}{\alpha}\right)/
\tanh\left(\frac{1}{\alpha}\right) & \alpha < 0 \\
\end{array}\right.
\end{equation}
so that for $\alpha=0$, the DF is isotropic, for $\alpha < 0$ 
it is radially anisotropic
and for $\alpha>0$ is is tangentially anisotropic. We shall take 
$\alpha = -0.5$, since the halo appears to be radially anisotropic. 

For a singular isothermal sphere
\begin{equation}
\rho(r) = \frac{V_c^2}{4 \pi G r^2}, \qquad \phi(r) = V_c^2 \ln \frac{r}{r_s}
\end{equation}
The corresponding DF is 
\begin{equation}
\label{gaia_eq:DF_E}
g(E) = \frac{\e}{16 \pi^2 G V_c \kappa} \exp\left[-\frac{2E}{V_c^2}\right],
\end{equation}
(Gerhard 1991) where
\begin{equation}
\kappa = \int_0^\infty \d u \,{\rm e}^{-u} \int_0^{\eta_{\rm max}} 
h_\alpha(\eta)
\frac{\eta \d \eta}{\sqrt{\eta^2_{\rm max} - \eta^2}}.
\end{equation}
Here $\eta_{\rm max} = \sqrt{2 {\rm e}} \sqrt{u} {\rm e}^{-u}$, with $u
= (E - \phi)/V_c^2$.  Since the density profile may be derived from
the initial distribution function as
\begin{equation}
\rho(r) = \frac{4 \pi}{r} \int_{\phi(r)}^{\infty} \d E g(E) L_c(E) 
\int_0^{\eta_{\rm max}} h_{a}(\eta) \frac{\eta \d \eta}{\sqrt{\eta^2_{\rm max}
- \eta^2}},
\end{equation}
the joint probability distribution of $E$ and $\eta$ at a given radius
$r_p$ is
\[ P(E,\eta) = \frac{4 \pi}{r_p \rho(r_p)} g(E) L_c(E) \frac{\eta h(\eta)}{
\sqrt{\eta^2_{\rm max}- \eta^2}},\] 
(van der Marel, Sigurdsson \& Hernquist 1997). Using Eq.(\ref{gaia_eq:DF_E}),
we find that the normalised cumulative probability distribution of $E$
is
\begin{equation}
\label{gaia_eq:cumul_DF_E}
\hat{P}(< E) = 1 - \exp\left[-\frac{E-\phi}{V_c^2}\right].
\end{equation}

We may derive the initial positions of the satellites by assuming the
profile given in Eq.(\ref{gaia_eq:rho_r}), and using, from
Eq.(\ref{gaia_eq:cumul_DF_E}), the energy as $E = \phi(r_p) - V_c^2 \ln (1
- \cal{R})$, with $\cal{R}$ a uniform random variable. With the energy
(or ${\cal{R}}$) we can compute $\eta_{\rm max} = \sqrt{2 {\rm e}}
\sqrt{\gamma} (1 - {\cal{R}})^{\gamma} 
\sqrt{-\ln (1-{\cal{R}})}$, 
for $\gamma = 1/2.5$
to mimic the stellar halo power law. Using $\eta_{\rm max}$ and the probability
distribution for $\eta$ we may derive the non-circularity of the orbits, and
in this way fully determine the phase-space initial position of a satellite.

\subsubsection{Internal properties of the satellites}

For the satellites we assume they initially have King profiles, as do
most of the satellites in the Local Group. Their present day total
luminosity is fixed to be that of the stellar halo. The luminosity of
each satellite is drawn from a Gaussian distribution with mean $2.5
\times 10^7 \sL$ and dispersion $10^7 \sL$, thus reproducing the
characteristic luminosities of Local Group dwarf spheroidals. The
initial number of satellites is thus set to be 33.  We assume that the
satellites follow the scaling relations (Burstein et al. 1997)
\begin{eqnarray*}
\log L &=& 5.35 + 1.80 \log \sigma_v, \\
\log R &=& -0.82 + 0.51 \log \sigma_v. 
\end{eqnarray*}
These relations allow us to derive from the luminosity $L$, the 
core radius $R$, and the central velocity dispersion $\sigma_v$.  Our
King models have a concentration parameter $c = \log{r_t/r_K} \sim
0.72$, where $r_t$ and $r_K$ are the tidal and King radii
respectively. The initial mass of the satellite is now also fully
determined.  

\vspace*{0.5cm}
\subsection{Galactic potentials}
\label{gaia_subsec:galaxy}

We will consider two different Galactic potentials.  In both cases,
our Galaxy has three components: a dark halo, a disk and a bulge, but we
take different functional forms for the potential. In
Model I, we take a dark logarithmic halo
\begin{equation}
\label{gaia_eq:halo}
\Phi_{\rm halo} = v^2_{\rm halo} \ln (r^2 + d^2),
\end{equation}
a Miyamoto--Nagai disk
\begin{equation}
\label{gaia_eq:disk}
\Phi_{\rm disk} = - \frac{G M_{\rm disk}}{\sqrt{R^2 + (a + 
\sqrt{z^2 + b^2})^2}},
\end{equation}
and a spherical Hernquist bulge
\begin{equation}
\label{gaia_eq:bulge}
\Phi_{\rm bulge} = - \frac{G M_{\rm bulge}}{r + c}, 
\end{equation}
where $d$=12 kpc and 
$v_{\rm halo}$~=~131.5 $\kms $; 
$M_{\rm disk} = 10^{11} \sm $, $a$ = 6.5 kpc 
and $b$~=~0.26 kpc; $M_{\rm bulge} = 3.4 \times 10^{10} \sm$ 
and $c$~=~0.7 kpc. This choice of parameters gives a flat rotation curve
with an asymptotic circular velocity of 186 $\kms$. 

\begin{figure}
\flushleft{\psfig{figure=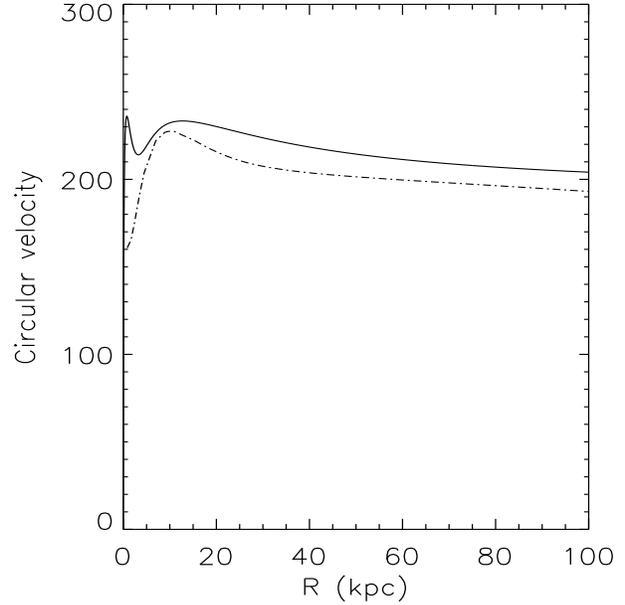,height=8cm,width=8.cm}}
\caption{The circular velocity profile as a function of distance from
the Galactic centre. The solid curve represents the potential used in
the simulations. The dashed curve corresponds to the alternative
potential.}\label{gaia_fig:circ_vel}
\end{figure}

In Model II, we represent the disk
density profile with a double exponential (Quinn \& Goodman 1986)
$$\rho_D(R,z)=\frac{M_D}{4 \pi R_D^2 z_o} \e^{-R/R_D} \e^{-\beta |z|}, $$
where $R_D = 3.5$ kpc is the disk scale length, $z_o$ is its scale height, 
$\beta = 1/z_o$ and $M_D = 5.5 \times 10^{10} \sm$ the total
disk mass. The associated potential is
\begin{eqnarray}
\label{gaia_eq:new_disk}
\Phi_D(R,z) &=& - \frac{GM_D}{R_D^3}\times \nonumber \\
&& \!\!\!\!\!\!\!\!\!\!\!\!\int_{0}^{\infty} \frac{\d k k J_0(kR)}
{k^2 + 1/R_D^2} \frac{\beta^2}{\beta^2 - k^2} \left\{\frac{\e^{-k|z|}}{k} -
\frac{\e^{-\beta|z|}}{\beta}\right\}.
\end{eqnarray}
For the halo we choose (Hernquist 1993)
$$ \rho_h(r) = \frac{M_h}{2 \pi^{3/2}}\frac{\alpha_q}{r_c}
\frac{\e^{-r^2/r_c^2}}{r^2 + \gamma_q^2}, $$ 
where $M_h = 1.5 \times 10^{12} \sm$, 
$\alpha_q = \left(1 - \sqrt{\pi}\,q \e^{q^2} (1 - {\rm
Erf}[q])\right)^{-1}$ with $q = \gamma_q/r_c$ and $r_c = 200$ kpc is the cutoff
radius. The corresponding potential is
\begin{equation}
\label{gaia_eq:new_halo}
\Phi_h(r) = - \frac{G M_h(r)}{r} + \frac{G M_h}{\sqrt{\pi} r_c} {\rm Ei}\left[
-\left(\frac{r}{r_c}\right)^2 - q^2\right],
\end{equation}
where
\begin{equation}
M_h(r) = \frac{2 M_h \alpha_q}{\sqrt{\pi}} 
\int_{0}^{r/r_c} \frac{x^2 \e^{-x^2}}
{x^2 + q^2} \d x
\end{equation}
and Ei$(x)$ is the exponential integral (e.g. Gradshteyn \& Ryzhik 1965). 
For the bulge we use Eq.(\ref{gaia_eq:bulge}) but we take $M_b = 1.1 
\times 10^{10}$ and $c = 0.525$ kpc (following Vel\'azquez \& White 1995). 
Figure~\ref{gaia_fig:circ_vel} shows the circular
velocity curves produced by each of the two potentials.

\subsection{Numerical methods}

In our numerical simulations, we use the potential of Model~I for our Galaxy. 
We represent the satellite galaxy by a collection of $10^5$ 
particles and model their self-gravity 
by a multipole expansion of the internal potential to fourth order 
\cite{W83,Zaritsky_White}. 
This type of code has the advantage that
a large number of particles can be followed in a relatively small
amount of computer time. In this quadrupole expansion, higher than monopole 
terms are softened more strongly. We choose
$\epsilon_1 \sim 0.2 - 0.25 R$ for the monopole term 
($R$ is the core radius of the system) and 
$\epsilon_2 = 2 \epsilon_1$ for dipole and higher terms 
and for the centre of expansion. The centre of expansion
is a particle which, in practice, follows
the density maximum of the satellite closely at all times.  

\begin{figure}
\flushleft{\psfig{figure=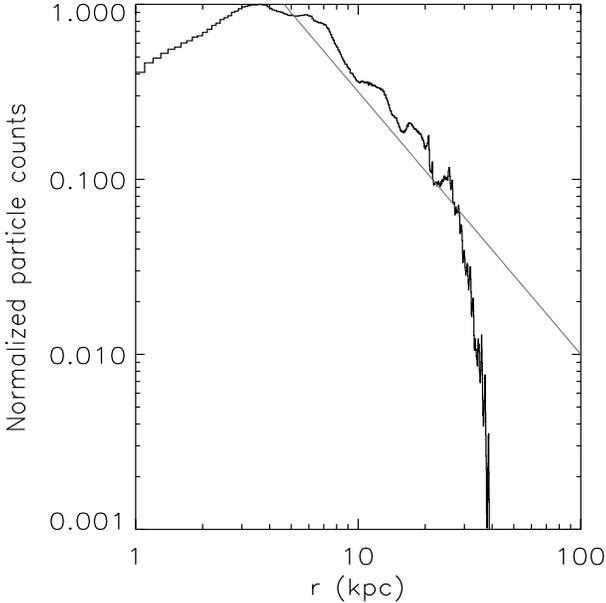,height=8cm,width=8.cm}}
\caption{Number counts profile $N(r) = r^2 \rho(r)$ for the simulated
stellar halo resulting from the superposition of 33 disrupted
satellite galaxies, after 12 Gyr of evolution. The straight line
represents the expected $r^{-1.5}$ law, arbitrarily
shifted.}\label{gaia_fig:profile}
\end{figure}

After letting our satellite relax in isolation, we integrate each
simulation for $\sim 12$ Gyr. In Figure \ref{gaia_fig:profile} we show
the final particle counts in radial bins $N(r) = r^2 \rho(r)$ as a
function of distance from the Galactic centre resulting from the
superposition of all our experiments. For guidance, we also plot the
expected $r^{-1.5}$, arbitrarily shifted. We see that within the range
of 3 to 30 kpc, our simulations follow relatively well the
profile. Outside this range we see a sharp drop, due to the fact that
we are (intentionally) not populating the outer halo. Since the
properties of the inner stellar halo are not so well constrained, we
do not worry about the fact that we find a shallower slope in the
inner few kiloparsecs (this is also the result of our initial
conditions). More important is the fact that our simulations can
reproduce very well the regime where the astrometric missions promise
to give accurate six dimensional phase-space information.

\subsection{Generating catalogues of halo stars}
\label{gaia_subsec:catalogs}

To generate an artificial catalogue for the Galaxy we assume that each
particle in our simulations represents a  giant star of absolute
magnitude $M_V = 1$. The total number of particles in our simulations
corresponds well to the expected number of giant stars in the Galactic
halo (based on the luminosity function, derived for the age and
metallicity characteristic of halo stars).  We prefer to take only
giant stars at this stage because they are bright enough to be easily
observable from the Sun. We need to determine a limiting magnitude
$V_{\rm lim}$ for our ``artificial'' catalogue, which we define so
that all giant stars brighter than $V_{\rm lim}$ have accurate full
6-dimensional phase-space information. In the case of GAIA, we take 
$V_{\rm lim} = 15$, a limit set by the accuracy in the radial velocity. 
For FAME, $V_{\rm lim} = 12.5$ as all stars brighter than this magnitude
will have relative parallax errors $\sigma_\pi/\pi$ 
smaller than (or of the order of) 25\%. For DIVA, we take $V_{\rm lim} = 11$
for which $\sigma_\pi/\pi \sim 0.3$. Our GAIA, FAME and DIVA 
catalogues have 386144, 12497  and 1742 ``stars'' with $M_V = 1$ 
respectively.

The positions and velocities of each particle are first transformed
into the observables ($\alpha$, $\delta$, $\pi$) and ($\mu_\alpha$,
$\mu_\delta$, $v_r$); the expected observational ``errors" are then
added to the parallax, the radial velocity and the proper motion,
according to Table \ref{gaia_tab:accuracy}.  For GAIA the precision in
the radial velocity is taken to be 5 $\kms$ for $V < 14$, and to vary
like $\sigma_v = 10 (V-14) + 10 \kms$ up to $V = 15$. Since FAME and
DIVA will not measure radial velocities on board, for these we
estimate the error $\sigma_v = 15 \kms$, as achievable from the ground
for such large samples. These ``observed" quantities are then
transformed back to ``observed" positions and velocities.  We repeat
this procedure 5 times to obtain 5 different realizations of the data.
\begin{table}
\caption{Estimated precision in parallax ($\sigma_\pi$, in $\mu$as)
and proper motion ($\sigma_\mu$, in $\mu$as ${\rm yr}^{-1}$) as a
function of $V$ magnitude. For FAME and DIVA we assume $\sigma_\pi =
\sigma_\mu$ (based on Horner (1999) and R\"oser (1998),
respectively). In the case of GAIA, the estimated precisions
correspond to a K3 III star with no reddening, and increase to 
0.2 mas at $V \sim 20$ (Gilmore et al. 1998).}\label{gaia_tab:accuracy}
\begin{center}
\begin{tabular}{ccccccccc}
& & 9 & 10 & 11 & 12 & 13 & 14 & 15 \\
\hline
GAIA & $\sigma_\pi$ & 3.65 & 3.65 & 3.65 & 3.65 & 4.83 & 7.05 & 10.8 \\
 & $\sigma_\mu$ & 2.74 & 2.74 & 2.74 & 2.74 & 3.62 & 5.28 & 8.10\\
FAME & $\sigma_\pi$ & 24  &    36 & 56   & 90    & 146 &  & \\
DIVA & $\sigma_\pi$ & 200  & 250  & 300  &       &     &   & \\
\hline 
\end{tabular}
\end{center}
\end{table}

\section{Finding disrupted galaxies}

\begin{figure*}
\begin{center}
\psfig{figure=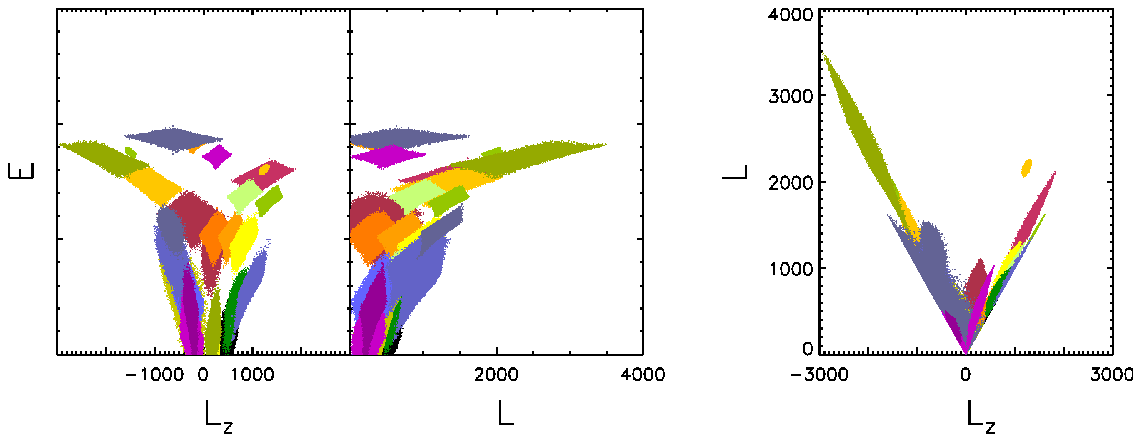,height=14cm,width=6.cm,angle=90}
\end{center}
\caption{Initial distribution of particles in the integrals of motion space. The different
colours represent different satellites.}
\label{gaia_fig:IC}
\begin{center}
\psfig{figure=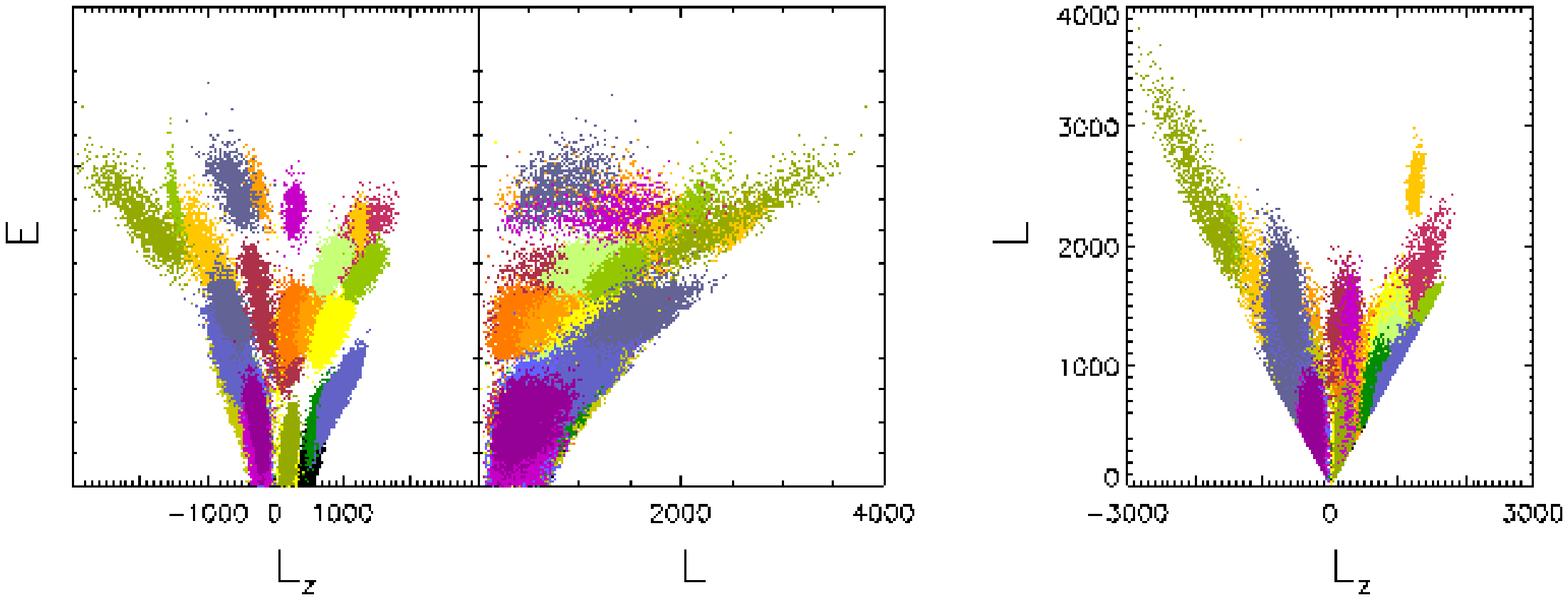,height=14cm,width=6.cm,angle=90}
\end{center}
\caption{Final distribution of particles in the integrals of motion
space after 12 Gyr, after convolution with the errors expected for
GAIA for the original potential. Here we include all particles
brighter than $V = 15$ (i.e. within roughly 6 kpc from the Sun).}
\label{gaia_fig:sub6kpc}
\vspace*{0.4cm}
\begin{minipage}[b]{0.46\linewidth}
\centering
\psfig{figure=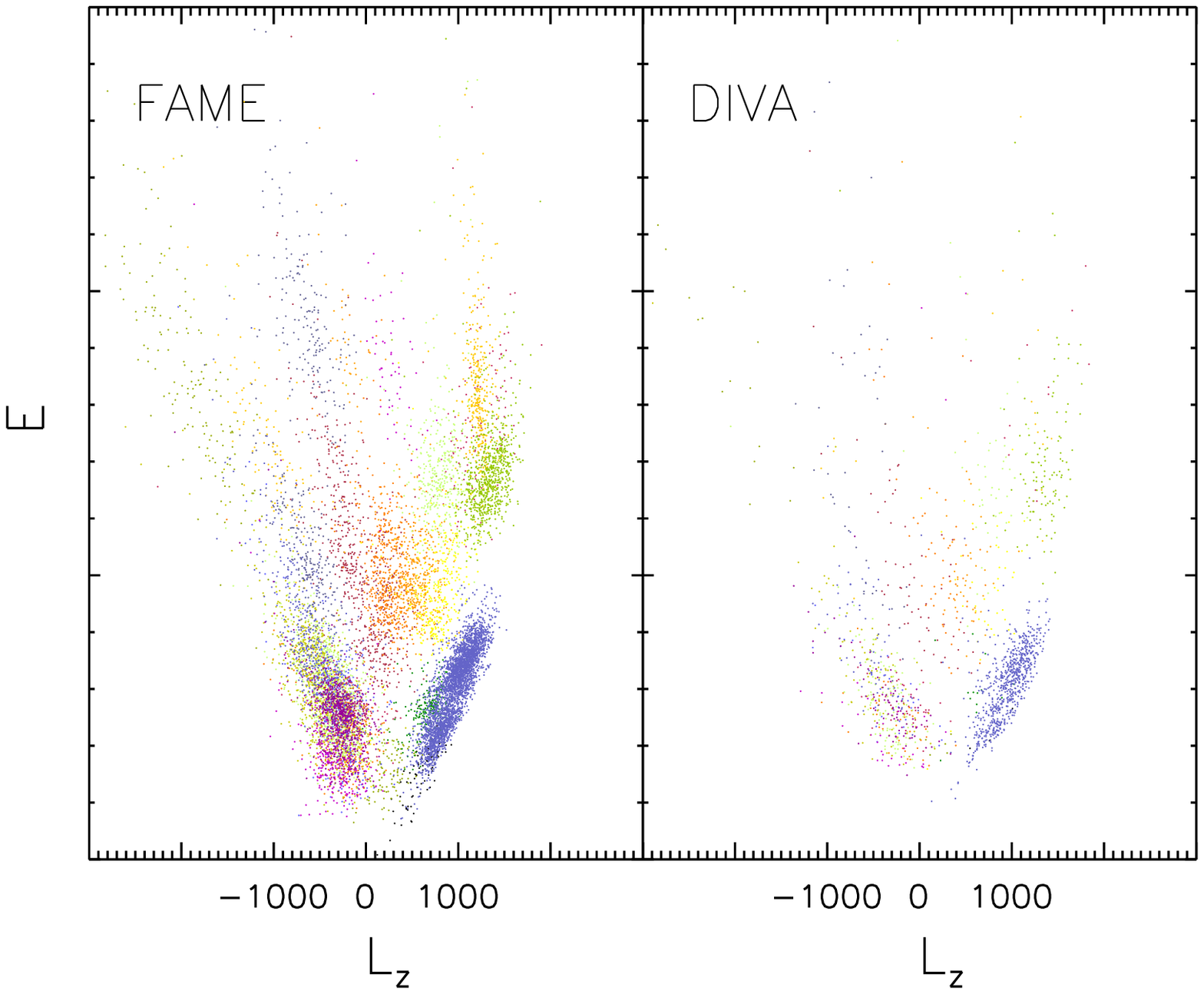,height=5.5cm,width=8cm}
\caption{Final distribution of particles in the $L_z$ -- $E$ space
after error convolution for FAME (left panel) and for DIVA (right
panel), with energies computed using the original potential. A
comparison to the left panel of Figs.~\ref{gaia_fig:IC} and
\ref{gaia_fig:sub6kpc} shows that the expected errors for these
missions tend to erase much of the substructure left in the integrals
of motion space.}
\label{gaia_fig:diva_fame}
\end{minipage}\hfill
\begin{minipage}[b]{0.46\linewidth} 
\centering
\vspace*{-0.2cm}
\psfig{figure=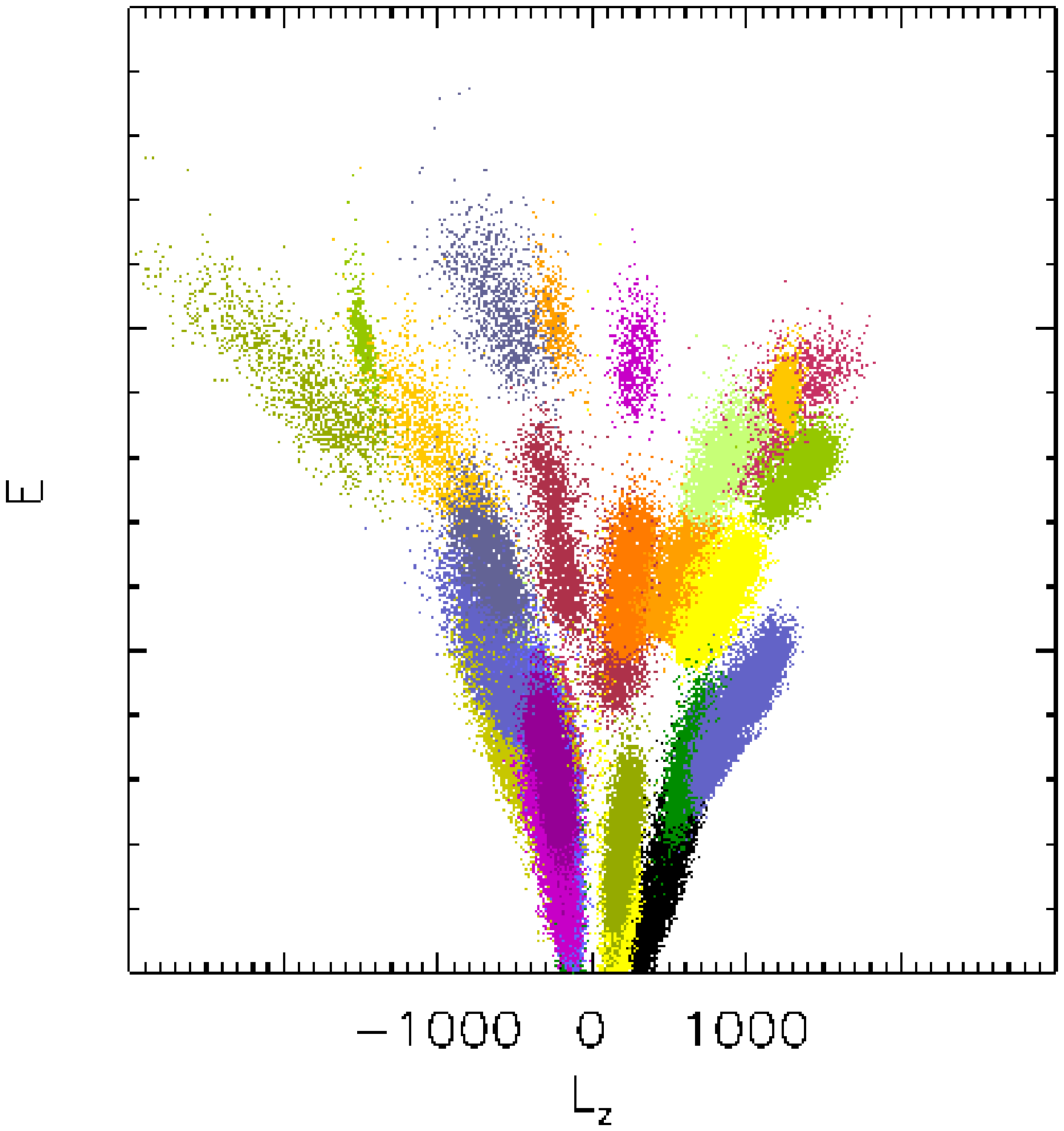,height=6.5cm,width=6.5cm}
\caption{Final distribution of particles in the $L_z$ -- $E$ 
space after GAIA error convolution for the alternative potential.
Compare to left panel of Fig.\ref{gaia_fig:sub6kpc}.}
\label{gaia_fig:sub6kpc_003.1}
\end{minipage}
\end{figure*}

\subsection{Integrals of motion space}

Our satellites disrupt relatively quickly, in only a few pericentric
passages. Therefore we may consider each of the 33 satellites as an
ensemble of particles with very similar integrals of motion (energy,
angular momentum).  As we show in Figure~\ref{gaia_fig:IC}, initially
satellites are both clumps in configuration and velocity space, as
they are in ($E$, $L$, $L_z$) space.  If these are conserved
quantities, or evolve only slightly, this initial clumping should be
present even after the system has phase-mixed completely. Thus the
space of integrals or adiabatic invariants seems to be the natural
space to look for the substructure produced by an accreted satellite.

There are a few issues we should address here before fully discussing
a method based on clumping in the integrals of motion space.  To
compute the energy of the particles (or stars that will be observed by
GAIA for example) we need to assume a Galactic potential. To determine
the success of such a method we need to understand how our lack of
knowledge on the precise form of the Galactic potential influences our
results. We shall therefore proceed in two steps. In the first step,
we take the same potential as that used in the simulations,
Model~I. In the final step we use the alternative potential introduced
in Sec.~\ref{gaia_subsec:galaxy}, our Model~II.  This last step, in
which we do not know the exact form of the Galactic potential but we
make a reasonable guess, is most likely to represent the real
situation.

Secondly, even though the total angular momentum is not fully
conserved for an axisymmetric potential (only $L_z$ is), it evolves
preserving a certain degree of coherence. The advantage of using the 
integrals of motion space is that the number of clumps detected in
this way will represent well the total number of accretion/merging
events, since unlike other methods which are only local, it singles
out all the stars from a given accreted object, independently of how
different their phases and velocities might be. We choose to make use
of all three integrals to reduce the chances of overlap amongst
different lumps, since this probability clearly depends on the
dimensionality of the space.

The analogue of Figure~\ref{gaia_fig:IC} for particles ``brighter than
15th magnitude'' (roughly within 6 kpc from the Sun) in the
simulations, after 12 Gyr of evolution and for the original potential,
shows that, even though there is some degree of evolution, clumping
remains in the integrals of motion space.  In Figure
\ref{gaia_fig:sub6kpc} we plot the integrals of motion space for one
realization of the GAIA catalogue, i.e. after error convolution. A
number of substructures are clearly visible, many of which can be
directly related to the initial distribution, even with the GAIA
observational uncertainties taken into account. This shows that the
expected observational errors for GAIA will not affect the chances of
detecting such substructures. In the case of FAME the situation is not
as good, as illustrated in the left panel of Figure
\ref{gaia_fig:diva_fame}, where the different lumps are less populated
(because of the magnitude limit) and considerably more smeared out
(because of the larger observational errors). For DIVA the clumping
has disappeared almost completely, as shown in the right panel of the
same figure.

Figure \ref{gaia_fig:sub6kpc_003.1} corresponds to the same
realization of the GAIA catalogue as used before, but with the
energies calculated using the case of the potential of Model
II. Clearly, even though the two considered potentials are different,
the substructure remains. The uncertainty in the precise form of the
Galactic potential therefore does not affect the likelihood of finding
disrupted satellites.

\vspace*{-0.5cm}
\subsection{Method: FOF in integrals of motion space}

We use a Friends-of-Friends (FOF) algorithm to find clumps
in the integrals of motion space. This method has been used frequently to
find bound halos in cosmological $N$-body simulations. The basic idea is
that all particle pairs separated by less than a fraction $\ell$ of the
mean interparticle distance are linked. Disjoint sets of
connected particles are then identified as halos (Efstathiou et
al. 1988). These halos correspond approximately to the regions
interior to isodensity contours at an overdensity of 2/$\ell^3$.  This FOF
procedure allows a rapid identification of halos, and moreover, all
members of a given halo found for a particular value of $\ell$ are
members of the same halo in any list generated for a larger value of
$\ell$. In the case of cosmological simulations, the linking distance is
defined so that the mean density of a halo is about 200 times the
density of the Universe at the time of identification.

In our case, it is less clear how we should define the interparticle
distance, or the linking length. Because the energy and angular
momentum have, by definition, different scales, it seems natural to
try to reduce everything to the same scale, or equivalently, to use
instead of spheres an ellipsoidal configuration.  Even though the
angular momentum and its $z$-component have the same scale, lumps are
generally elongated in the $L$-direction with a 2:1 ratio, as can be
seen from Figure~\ref{gaia_fig:IC}.  We therefore search for lumps
whose characteristic size would be defined as:
\[ \Delta L \sim 2 \Delta L_z, 
\qquad \frac{\Delta E} {(\kms)^{2}} \sim 20 \frac{\Delta L_z}{{\rm kpc} \kms}, \]
where now $\Delta L_z$ would be related to the linking length. This implies
that we re-scale the variables according to
\[E \rightarrow E/20, \qquad L \rightarrow L/2, \qquad L_z \rightarrow L_z.\]
The factor $20$ in the energy scaling may be derived (heuristically)
from the fact that the typical energy range in the Solar neighbourhood
is $1.6 \times 10^5 (\kms)^2$, whereas the range of $L_z$ 
is $8000$ kpc $\kms$ (from $-4000$ to $4000$ kpc
$\kms$).

We will apply the FOF algorithm for two different linking lengths, to
allow for different characteristic sizes of the halos and resolutions
in the algorithm.  Note that there are particular regions in this
space which are occupied by more than one satellite, even in this
3-dimensional space (this is even worse if only the $L_z-L$ plane is
used), so that not each of the lumps found may correspond to only one
satellite, but may have contributions of a few.

\vspace*{-0.4cm}
\subsection{Results}

\begin{figure*}
\psfig{figure=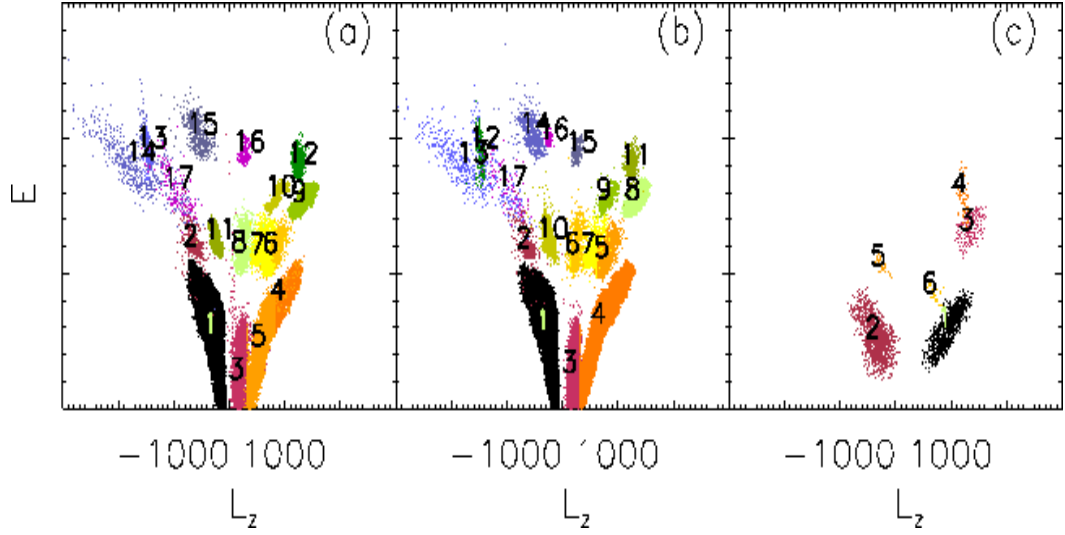,height=14cm,width=7.cm,angle=90}
\caption{Lumps detected with our FOF algorithm. In panel (a) we show
the final group catalogue for the original potential used in the
simulations after convolving with the observational errors expected
for GAIA. Panel (b) corresponds to our alternative potential and also
to the GAIA catalogue. In both cases the recovery rate is about
50\%. Panel (c) shows the lumps recovered by our FOF applied to the
FAME catalogue generated as described in the text and for energies
computed with the original potential. Compare to
Fig.\ref{gaia_fig:sub6kpc} in the case of GAIA and to the left panel
in Fig.\ref{gaia_fig:diva_fame} for FAME.}
\label{gaia_fig:FoF_000_6kpc.1}
\end{figure*}

We apply the FOF algorithm to our GAIA catalogue, including error
convolution for all particles brighter than 15th magnitude, and using
the original potential. We take two different values for the FOF
linking length: $\ell = 16$ and $\ell = 30$, where $\Delta L_z = 5
\ell$. We consider groups with
at least 500 ``stars''. We combine the two group catalogues to obtain
a new group catalogue which contains all lumps detected. If some
particles are found to belong to two different clumps (one from each
catalogue) we keep the lump which has the smallest size.  We now
iterate one more time on the catalogue defined by the particles that
do not belong to any of the lumps found by our FOF, again for the two
values of $\ell$ and with a minimum of 250 ``stars''. Some of the
newly found lumps can be related to those previously detected, and
some others are found to resolve some of the largest lumps in our
initial group catalogue. In the left panel of Figure
\ref{gaia_fig:FoF_000_6kpc.1} we show the distribution of energy $E$
and $L_z$ for our final group catalogue.

\begin{figure*}
\psfig{figure=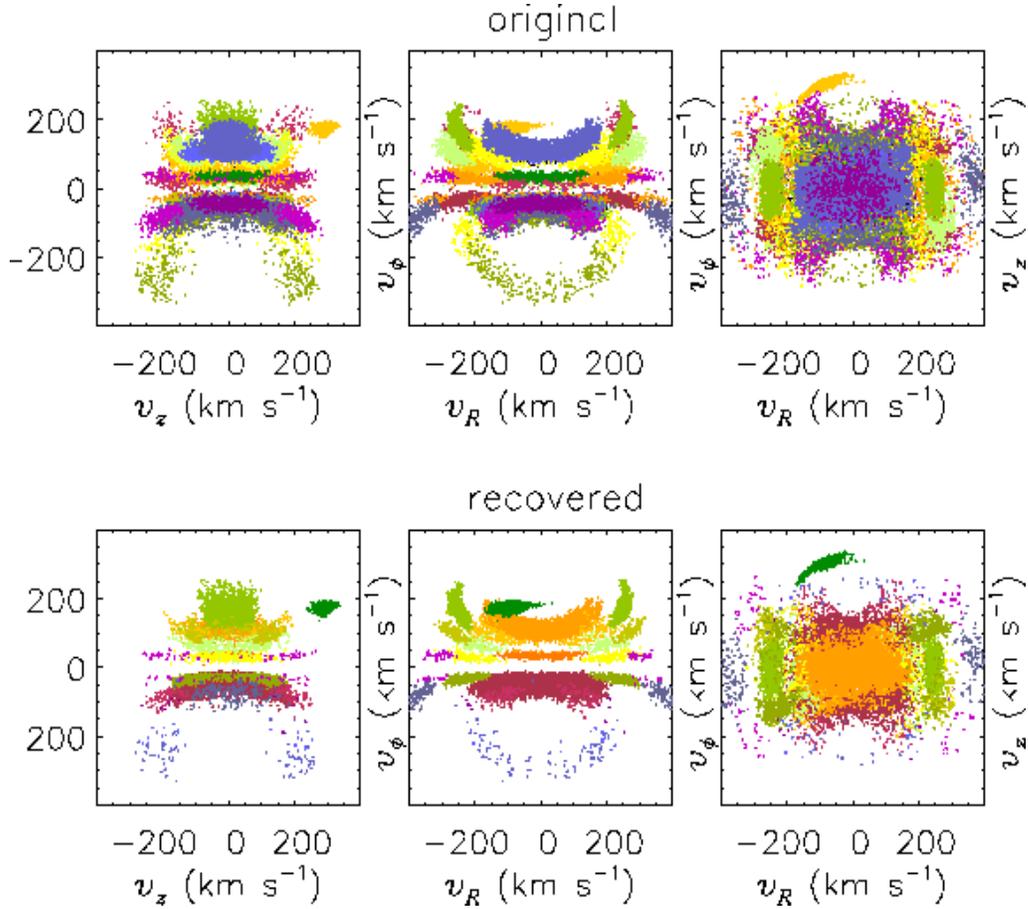,height=12.cm,width=13.5cm}
\caption{The velocity space distribution for particles in a cubic
volume of 2 kpc on a side centered on the Sun, and for one realization
of the GAIA catalogue. In the upper panels different colours indicate particles
associated with different satellites (using the same colour coding
as in Figure \ref{gaia_fig:IC}).  In the lower panels, the colours are
used to show particles associated to the lumps recovered by our FOF
algorithm applied to the GAIA catalogue in the case of the alternative
Galactic potential. (Here the colour coding corresponds to that used in
panel (b) of Figure \ref{gaia_fig:FoF_000_6kpc.1}.)}
\label{gaia_fig:vel}
\end{figure*}

We find 17 different groups with this method. Not all the groups may
be associated exclusively with one of our original satellites. As can
be seen from Fig.\ref{gaia_fig:IC}, there is quite a bit of
superposition in this three--dimensional space, and so not all the
original satellites can be recovered, or equivalently, not all lumps
can be resolved with just two iterations. If we analyse how the
particles in the different lumps can be related to particles in the
initial satellites we find that, out of the 17 groups discovered, 14
can be associated almost uniquely to one satellite\footnote{We say
that a group is almost uniquely associated to one satellite if more
than 70\% of the particles in the group belong to only that
satellite.}. This means that our simple method is capable of finding
more than 40\% of all the satellites that were accreted by our
``Galaxy''.

Similarly, we apply the FOF algorithm to the same GAIA catalogue but
now compute the energy $E$ of the particles with the alternative
potential.  In this case we again find 17 different lumps (after two
iterations, and combining the results of the two different values of
the linking length). Of these 17 groups, 14 can be uniquely associated
to one satellite. This is shown in the central panel of Figure
\ref{gaia_fig:FoF_000_6kpc.1}.  Our method is thus quite successful in
identifying disrupted satellite galaxies in integrals of motion
space, even when we only have a guess for the Galactic potential and
when the observational uncertainties are taken into account. Further
uncertainties such as the distance from the Sun to the Galactic centre
and the velocity of the Local Standard of Rest are also negligible. 

When we apply the same method for the original potential on the FAME
catalogue we are able to find 6 different groups. For 5 of these a
unique correspondence with an accreted satellite exists, as shown in
the rightmost panel of Figure \ref{gaia_fig:FoF_000_6kpc.1}. In the
case of DIVA we find only 1 group (with at least 20 particles), which
can be easily identified visually from the right panel in Figure
\ref{gaia_fig:diva_fame}. In the cases of DIVA and FAME we used
slightly larger linking lengths to take into account the smearing out
of the lumps caused by the larger observational uncertainties. Because
the samples are also smaller (because of the limiting magnitude), we
consider groups with at least 50 particles in the case of FAME, and 20
particles for DIVA. 

\section{Clumpiness in the kinematics of halo stars}

In Figure \ref{gaia_fig:vel} we show the velocities of all the
particles contained in a volume of 2 kpc on a side for one realization
of the GAIA catalogue. There is considerable substructure, which is
visible thanks to the great precision that GAIA will achieve. From the
upper panels it is clear that, as discussed in the introduction,
distinguishing the satellites that gave rise to each one of the
different moving groups is a non-trivial task in this space. In the
lower panels we have coloured the different contributions from the 14
groups detected by our FOF algorithm in the case of the alternative
potential. A comparison between upper and lower panels also shows how
successful our method is.

The kinematically cold streams visible in Figure \ref{gaia_fig:vel}
remain as coherent structures for longer than a Hubble time. This is
true even when mergers, rather than simple satellite accretion, are
dominant (Helmi, White \& Springel 2000). The clumpiness in the
kinematics of halo stars should thus be a distinct feature of the
hierarchical formation of our Galaxy.  It is therefore also
interesting to determine the degree of the clumpiness and whether it
will be measurable with future astrometric missions.  We will
determine this clumpiness using the two--point correlation function
$\xi$ in velocity space for a sphere of 1 kpc radius around the
Sun. We estimate $\xi$ from
\begin{equation}
\xi = \frac{\langle DD \rangle \langle RR \rangle}{\langle DR \rangle^2} - 1 
\end{equation}
(e.g. Hamilton 1993) where $\langle DD \rangle$ is
the normalised number of pairs of particles with
velocities in a given velocity range (or bin), i.e.
\begin{equation}
\langle DD \rangle = \!\!\!\frac{\sum \mbox{\small 
pairs of particles $i,j$ with }  
v < |{\bf v}_i \! - \! {\bf v}_j| < v + \Delta}{N_D (N_D - 1)}
\end{equation}
and where $N_D$ is the number of particles in the sphere.  $\langle RR
 \rangle$ is defined analogously but for $N_R$ random points. 
The random variates are drawn from 
 a trivariate Gaussian distribution determined from the ``data'' in the
 principal axes velocity frame. Here we take $N_R = 10 N_D$. Finally 
$\langle DR \rangle$ are the normalised counts for ``data''--random pairs. 
We estimate the uncertainty  from 
\begin{equation}
\Delta_\xi = (1 + \xi) \sqrt{\frac{2}{N_D (N_D - 1) \langle DD
\rangle}}.
\end{equation}

\begin{figure*}
\psfig{figure=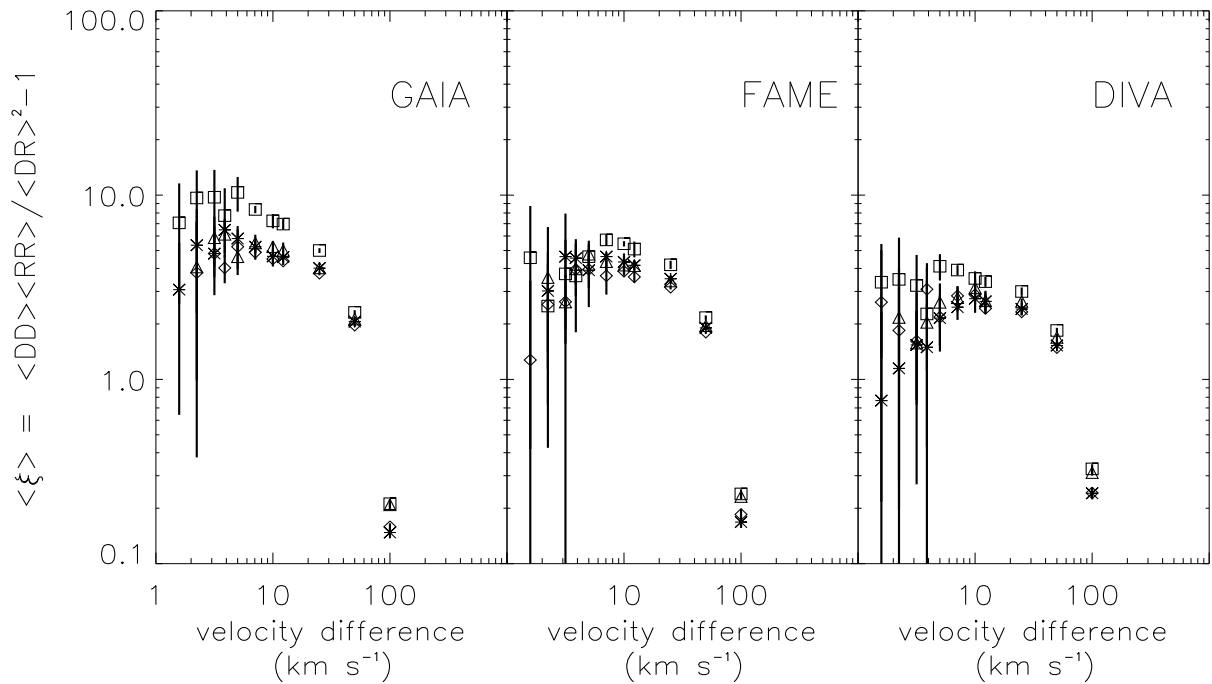,height=16cm,width=9.cm,angle=90}
\caption{The two-point correlation function for ``giant stars'' inside
spheres of 1 kpc radius around the Sun (defined as 8 kpc from the
Galactic centre on the Galactic disk) computed as the weighted average
over five realizations of the DIVA,
FAME and GAIA catalogues. The different
symbols correspond to $\xi$ measured inside spheres at different
locations of the ``Sun'' on the Solar circle.}\label{gaia_fig:xi}
\end{figure*}

If the sample contains kinematically cold streams, we should find an
excess of pairs in the bins corresponding to small velocity
differences, i.e. the correlation function should be significantly
different from zero (which corresponds to the absence of
correlations). We proceed by measuring $\xi$ for our GAIA, FAME and
DIVA catalogues including error convolution as described in
Sec.~\ref{gaia_subsec:catalogs}.  We also vary the position of the 1
kpc sphere around the ``Sun'', keeping the same distance from the
``Galactic centre''. That is, we place the Sun at $(x,y) =
\{(8,0),(0,-8),(-8,0),(0,8) {\rm kpc} \}$ and $z = 0$. This allows us
to account for the natural variations one may have from volume to
volume.  We then make five realizations for each ``Sun'' position for
each catalogue. In Figure \ref{gaia_fig:xi} we show the correlation
function obtained by averaging over all the realizations, for each 1
kpc sphere and for each catalogue. The average $\xi$ for each volume
is the weighted mean, where the weights are given by $1/\Delta_\xi^2$,
and the error bars indicate the (weighted) dispersion around the
(weighted) mean. We find an excess of pairs of stars with similar
motions, the signature indicating the presence of cold streams as
expected for a stellar halo built by disrupted satellites. Note that
it will even be possible to determine that the halo is not a smooth
distribution in the Solar neighbourhood even with velocity errors of
the order of 20 $\kms$ such as those expected for DIVA 
for a star with $M_V = 1$ at 500 pc from the Sun. 

\section{Discussion}

We simulated the entire stellar halo of the Galaxy starting from disrupted
satellite galaxies, and ``observed'' it with the next generation of
astrometric satellites (DIVA, FAME and GAIA).  We analysed the
observations with the aim of recovering the different accretion events
our ``Galaxy'' experienced over its lifetime. We used a FOF algorithm
to find clumps in the integrals of motion space, which we expected
would correspond to the disrupted satellites. Our integrals of motion
space is defined by energy $E$, total angular momentum $L$ and its
$z$-component $L_z$, even though strictly speaking these are not fully
conserved quantities (because of interaction of the stars while still
bound to the satellite, and because of the axisymmetry of our Galaxy).
We have shown that the initial clumping in this space is maintained to
a great extent even after 12 Gyr of evolution.

After using our FOF algorithm we find that we can only recover a
couple of accreted satellites (in our analysis just one) for the DIVA
catalogue, whereas for FAME we recover about 15\% of all
satellites. In both cases we assume that the astrometry is
complemented by ground based radial velocity measurements. The
situation is significantly different in the case of the GAIA
catalogue, for which we recover almost half of all disrupted
satellites with this simple algorithm. The improvement generally lies
in the larger volume for which full 6-D information is available, in
particular when comparing FAME and GAIA.  The use of 6-D information
appears to be essential to recover all the events, as there is a large
fraction of phase-space where these are superposed. This is
particularly clear from Figure \ref{gaia_fig:sub6kpc} (rightmost
panel), where angular momentum alone cannot be used to distinguish the
different satellites. Whereas by eye inspection in the ($E$, $L$,
$L_z$) space we may recover five or six events, for the space ($L$,
$L_z$) this is reduced to one or two events. Our results are unlikely
to be strongly dependent on the particular choice of the luminosity
distribution of the disrupted satellites. This is because a large 
number of small satellites occupy basically the same phase-space
volume as a small number of large ones.

The evolution of the Galactic potential may be the most crucial
simplification in our analysis.  In hierarchical cosmologies the
number of objects that form a galaxy like our own is in the range of
$5- 20$, with comparable masses.  The process of formation is likely
to be very violent and the potential is surely not static, quite
probably not axisymmetric, and therefore the initial clumping of the
system may not be reflected in clumping in our defined integrals of
motion space. However, if this happened during the first few Gyrs, any
object infalling later, ought to have perceived a fairly static (or
adiabatically changing) Galaxy, and then our method would still be
useful.  Indeed, some preliminary analysis of the formation of a halo
in a $\Lambda$CDM cosmology indicates that particles from different
satellites may be recovered as lumps in this space (Helmi et
al. 2000), though the structure is less evident than in the plots
shown here, where even by simple eye inspection one may recover about
1/5 of all satellites.

What will anyway remain as signatures of the merger history of our
Galaxy will be the kinematically cold streams originating in disrupted
halos.  An interesting observational test is the comparison of the
kinematics of a smooth, possibly Gaussian, distribution (which may be
expected in the case of a monolithic collapse) to the kinematics
observed in the stellar halo built by disrupted satellites. Our
analysis of the correlation function in velocity space indicates the
presence of a larger number of streams with very small velocity
dispersions in a sphere of 1 kpc radius around the Sun. This test will
be feasible even for DIVA. The key to the success of this test lies in
the complete and large sample of stars with 3-D velocities which will
be available.

In this paper we have focused on determining the merger history of the
Milky Way, rather than the precise form of the Galactic potential or
to what extent it may have varied.  However, these are key questions
that will be solved very likely by SIM and GAIA (e.g. Johnston et al.
1999).  We may add here that after finding the different satellites we
will be able to determine the conditions and characteristics of
objects that fell onto the Milky Way more than 10 Gyr ago. 

\section*{Acknowledgments}
We wish to thank  Volker Springel for his FOF algorithm, Simon White
for many useful discussions, and Anthony Brown for comments on an earlier
version of this manuscript.

\label{lastpage}

\end{document}